\catcode`\@=12 
\documentstyle[12pt]{article} 
\newcommand{\be}{\begin{equation}}
\newcommand{\en}{\end{equation}}
\newcommand{\ba}{\begin{eqnarray}}
\newcommand{\ea}{\end{eqnarray}}
\newcommand{\bea}{\begin{eqnarray*}}
\newcommand{\eea}{\end{eqnarray*}}

\newcommand{\no}{\noindent}
\newcommand{\vs}{\vspace}

\newtheorem{Th}{Theorem}

\newtheorem{Pro}{Proposition}
\title{{\bf Infinite dimensional SRB measures}}

\author{J.Bricmont\thanks{Supported by EC grant CHRX-CT93-0411}\\ UCL,
Physique Th\'eorique, B-1348, Louvain-la-Neuve,
Belgium\\bricmont@fyma.ucl.ac.be\and
A.Kupiainen\thanks{Supported by NSF grant DMS-9205296 and EC grant
CHRX-CT93-0411} \\
Helsinki University, Department of Mathematics,\\ Helsinki 00014,
Finland\\ajkupiai@cc.helsinki.fi}

\date{}

\begin{document}

\maketitle \begin{abstract}
We review the basic steps leading to the construction of a Sinai-Ruelle-Bowen
(SRB) measure
for an infinite lattice of weakly coupled expanding circle maps, and we show
that this
measure has exponential decay of space-time correlations. First, using the
Perron-Frobenius
operator, one connects the dynamical system of coupled maps on a
$d$-dimensional lattice to   an equilibrium
statistical mechanical model on a lattice of dimension $d+1$. This lattice
model is, for weakly coupled maps, in a high-temperature phase, and we use a
general, but very elementary, method to prove exponential decay of correlations
at high temperatures.
 \end{abstract}

\section{Introduction.}
\setcounter{equation}{0}

One of the landmarks in the modern theory of dynamical
systems was the introduction of the
concept of Sinai-Ruelle-Bowen (SRB) measure \cite{Si,Si2,Ru,Bo,ER}.
This is, roughly speaking, a measure that is supported on
 an attractor and that describes the
statistics of the long-time behaviour of the orbits, for  almost every
initial condition (with respect
to Lebesgue measure) in the corresponding basin of attraction.
On the other hand, one expects the asymptotic behaviour
of solutions of a class of partial differential equations, defined on some
spatial domain, to be described by an attractor whose dimension increases to
infinity with the size of the domain \cite{Te}.
It seems therefore natural to try to extend the notion of SRB measure to
infinite dimensional dynamical systems.

 One often
replaces differential equations
 by iterated maps, i.e. recursions of the form:
$$ x(t+1) =
F( x(t)).
$$
Likewise, in order to simplify the analysis of
 partial differential equations,
one may discretize space
and time, and consider instead coupled map lattices.  One replaces the field
described by the partial differential equation by a set ${\bf x} =(x_i)_{i\in
{\bf Z}^d}$. The dynamics is defined
 by  a recursion of the form:
$$ x_i(t+1) =
F_i({\bf x}(t))
$$
 i.e. $x_i(t+1)$ is
determined by the values taken by
${\bf x}$ at time $t$ (usually on the sites in a
neighborhood of $i$) \cite{Ka,Ka2}.

In the study of SRB measures, a major role is played by the thermodynamic
formalism: these measures are realized as equilibrium states for a ``spin
system", where the spin configurations correspond to a coding (via a Markov
partition) of the phase space of the dynamical system. But for smooth, finite
dimensional, dynamical systems, the corresponding spin system is always
one-dimensional, with exponentially decaying interactions. And these systems
do not undergo phase transitions, which then implies that the SRB measure is
unique.

When one considers coupled maps on a $d$-dimensional lattice, the
thermodynamic formalism relates them to a spin system  on
a  $d+1$-dimensional lattice, for which a phase transition is at least
possible. Some concrete models of such transitions have been suggested
\cite{MH,Po}.
Unfortunately, all rigorous studies of coupled maps have dealt, so far, with
the case of weakly coupled maps, where the corresponding spin system is its
high-temperature phase \cite{BS,PS,V,V2,V3,J,K,BK2,BK1}. However, it turns
out that the ``interactions" between the spins are somewhat unusual, and
even the high-temperature phase requires a careful analysis. The
investigation of a possible ``low temperature" phase is an open problem.

The goal of this paper is to provide a pedagogical introduction to the
methods used in \cite{BK1} to construct  the SRB measure for coupled
maps, and to analyze  its space-time decay properties. We believe that
the construction presented here is quite simple.

First of all, we give a unified presentation of high-temperature expansions
for spin systems.
At high temperatures, one generally expects a statistical mechanical system
to have three properties: a unique phase, exponentially small correlations
between
distant variables and analyticity of the thermodynamic functions with respect
to
parameters such as temperature, external field etc... We start by showing
how to prove the first two of those properties, starting with
the simple
example of the Ising model (using an argument that goes back to Fisher
\cite{Fi}), and then proceeding by successive  generalizations. Our method is
closely related to the high temperature expansions for disordered systems
developed in \cite{VKP}. Eventually, we show how to analyze a small, but
rather general, perturbation  of a lattice of one-dimensional
systems defined by subshifts of finite type. This result is slightly different
from those of \cite{BK1}.

Then we show in detail how to use the Perron-Frobenius operator to construct
a ``spin" representation of a dynamical system, for a {\it single} map.
We also explain how the time correlations of the dynamical system are
translated into spatial correlations for the corresponding spin system. The
construction of an SRB measure for the coupled maps follows then, in a
 rather straightforward way, from the combination of those two ingredients.

 \section{High-temperature expansions.}

\vspace{5mm}

Consider an Ising model on a $d$-dimensional lattice ${\bf Z}^d$: at each site
$i \in {\bf Z}^d$,
we have a variable $s_i= \pm 1$. Let $s_\Lambda=(s_i)_{i\in \Lambda}$, for
$\Lambda \subset{\bf Z}^d$.
In the simplest
case,
 we have the Hamiltonian
\be
{\cal H}(s_\Lambda)= -J \sum_{\langle ij \rangle} s_i s_j \label{1}
\en
with $J>0$, and the sum runs over nearest-neighbor ($n.n.$) pairs.
Consider the two
point correlation function $\langle s_0 s_k
\rangle_\Lambda$ where $0, k \in
\Lambda \subset {\bf Z}^d$
and the expectation $\langle \cdot \rangle_\Lambda$ is
taken with respect to the
distribution
\be
Z(\Lambda)^{-1} \exp (-\beta {\cal H}(s_\Lambda))
\label{2}
\en
where ${\cal H}(s_\Lambda)$ is defined as in (\ref{1}), but with
the sum restricted to
$\langle ij\rangle
\subset \Lambda$ (open boundary conditions), and $Z(\Lambda)$
normalizes (\ref{2}) into a
probability distribution. So,
\be
\langle s_0 s_k \rangle_\Lambda = \frac{\sum_{s_\Lambda}
 s_0s_k \exp
(-\beta {\cal H}(s_\Lambda))}{\sum_{s_\Lambda} \exp(-\beta
{\cal H}(s_\Lambda))}.
\label{3}
\en
We write
\be
\exp(-\beta {\cal H}(s_\Lambda)) = (\cosh \beta J)^{|B(\Lambda)|}
\prod_{\langle ij \rangle \subset \Lambda}
(1+ s_is_j \tanh \beta J)
\label{4}
\en
where $|B(\Lambda)|$ is the number of nearest-neighbor pairs in $\Lambda$.
Next, we insert (\ref{4}) in the numerator and the denominator
of (\ref{3}), cancel
the common factor $(\cosh \beta J)^{|B(\Lambda)|}$ and expand the product
over $\langle ij \rangle \subset \Lambda$. Using
 $\sum_{s_i=\pm 1} s_i = 0$, $\sum_{s_i=\pm 1} 1=2$, we get
\be
\langle s_0 s_k \rangle_\Lambda = \frac{\sum_{{\cal X}: \partial {\cal X}
=\{0,k\}}
(\tanh \beta J)^{|{\cal X}|}}{\sum_{{\cal X}:\partial {\cal X}
=\emptyset} (\tanh \beta J)^{|{\cal X}|}}
\label{5}
\en
where the sums run over sets of $n.n.$ pairs ${\cal X}$
and $\partial {\cal X}
$ is the set
of sites which appear an odd number of times in the
pairs of ${\cal X}$. Now, each ${\cal X}$
 with $\partial {\cal X}
= \{0,k\}$  contains a subset $P=\{\langle i_t,i_{t+1}\rangle \}^{n-1}_{t=0}$
where $i_0=0$, $i_n=k$ (otherwise, $\partial {\cal X} $ would contain points
other than $ 0$, $k$). By erasing loops in $P$, one can even choose $P$
to be a self-avoiding
path, i.e. such that all the $i_t$'s are different (but we shall not use this
last fact in any essential way below). So, each
${\cal X}$ in the numerator of (\ref{5}) can be written as  ${\cal X}=P
\cup {\cal X}'$ where $P$ is as above, and ${\cal X}'$ satisfies
$\partial {\cal X}'=\emptyset$. This decomposition of
${\cal X}$ is not necessarily unique, but,
given an arbitrary choice of a path
$P$ for each ${\cal X}$ in the  numerator of (\ref{5}), we may
write:
\ba
\sum_{{\cal X}:\partial {\cal X}=\{0,k\}} (\tanh \beta J)^{|{\cal X}|}
= \sum_{P:0\to k}
 (\tanh \beta J)^{|P|}
\sum_{ {\cal X}',\partial {\cal X}'=\emptyset}{^{\hspace{-5mm}P} } \;\;\;
(\tanh \beta J)^{|{\cal X}'|}
\label{6}
\ea
where $\sum^P$ runs over the sets of pairs such that $P$ is the chosen path
for
${\cal X}= P \cup {\cal X}'$. Clearly, since $\tanh \beta J \geq 0$,
the sum over ${\cal X}'$
in (\ref{6}) is
less than the unconstrained
sum over ${\cal X}$ in the denominator of (\ref{5}), and we get
\be
\langle s_0 s_k \rangle_\Lambda \leq \sum_{P:0 \to k}  (\tanh \beta J)^{|P|}
\label{7}
\en
Recall that $P=\{ \langle i_t,i_{t+1}\rangle \}^{n-1}_{t=0}$
where $i_0=0$, $i_n=k$
and all the $i_t$'s are different. Given $i_t$, we can
 choose at most $2d-1$
nearest-neighbor sites $i_{t+1} \neq i_{t-1}$ and, if
\be
(2d-1) \tanh \beta J < 1
\label{8}
\en
we have, since all $P$ in (\ref{7}) satisfy $|P| \geq |k|$
(with $|P|$ being
the number of pairs in $P$),
\be
\langle s_0
 s_k \rangle \leq \frac{2d}{2d-1}\sum_{n \geq |k|} ((2d-1)
\tanh \beta J)^n \leq C
\exp (-m |k|)
\label{9}
\en
for some $m>0$ and $C < \infty$ (the factor $2d$ comes from
the first step in $P$). This proves the exponential decay of the two-point
correlation function $\langle s_0
 s_k \rangle$, for $\beta$ small.

\vs{3mm}

\par\noindent {\bf Remark.} Here and below, we shall
denote by $C$ or $c$ a constant that
may vary from place to place.

\vspace*{3mm}

Let us now extend this argument in several directions. First, it is
instructive to abstract the argument and apply it to more general
 correlation functions.
Let $F$ depend on the variables $s_A = (s_i)_{i\in A}$ and
$G$ depend on $s_B = (s_i )_{i \in B}$, where
$A,B \subset \Lambda$. Then, for $\beta$ small,
\be
| \langle F G \rangle_\Lambda
- \langle F \rangle_\Lambda \langle G \rangle_\Lambda | \leq C ||F||
  \; ||G||
\exp (-md(A,B))
\label{10}
\en
for some $C < \infty$ (depending on $|A|$ and $|B|$)
and $m>0$, where $\| F \|$ is the sup norm
and $d(A,B)$ is the distance
between $A$ and $B$. The constants $C$ and $m$ are independent of
$\Lambda$.

To prove (\ref{10}), one introduces duplicate variables, i.e. one
associates to each lattice site a pair of
identically distributed random variables $(s^1_i, s^2_i)$
 whose common distribution  is given by (\ref{2}). Using this trick, one may
write \ba
&&2(\langle FG
\rangle_\Lambda - \langle F \rangle_\Lambda \langle
G \rangle_\Lambda)\nonumber \\
&=& Z(\Lambda)^{-2}
\sum_{(s^1_i,s^2_i)_{i\in \Lambda}}
(F(s^1_A)-F(s^2_A))(G(s^1_B)-G(s^2_B))
 \exp (-{\cal H}( s^1_\Lambda) - {\cal H}(s^2_\Lambda)).
\label{11}
\ea

Then, we use the following identity, which is analogous to (\ref{4}):
\be
\exp (\beta J (s^1_i s^1_j + s^2_i s^2_j)) = e^{-2\beta J}
(1+\exp (\beta J (s^1_i s^1_j + s^2_i s^2_j + 2))-1) \equiv
 e^{-2\beta J} (1+f_{ij})
\label{12}
\en
where, using $e^x -1\leq xe^x$, for $x\geq  0$, we have
\be
0\leq f_{ij} \leq 4\beta J  (1+f_{ij}).
\label{13}
\en
 Now we insert (\ref{12}) in the numerator and denominator of (\ref{11}),
cancel the
common factor $e^{-4\beta J|B(\Lambda)|}$, and
expand  $\prod_{\langle ij \rangle} (1+f_{ij})$ in the numerator;
 we obtain:
\be
\langle FG \rangle_\Lambda - \langle F \rangle_\Lambda \langle G
\rangle_\Lambda =
\frac{\sum_{\cal X} \langle \tilde F \tilde G f_{\cal X} \rangle^0}{\langle
\prod_{\langle ij \rangle \subset \Lambda }
(1+f_{ij})\rangle^0}
\label{14}
\en
where
$\tilde F = F(s^1_A) - F(s^2_A)$, $\tilde G = G(s^1_B) - G(s^2_B)$,
$f_{\cal X}=\prod_{\langle ij \rangle \in {\cal X}}
f_{ij}$, the sum
runs over sets ${\cal X}$ of $n.n.$ pairs and
 $\langle f \rangle^0$ means that we
sum $f$ over $s^1_i,s^2_i = \pm 1 , i \in \Lambda$.

It is easy to see that the terms in the numerator of (\ref{14})
for which ${\cal X}$
does not contain a path $P = \{ \langle i_t , i_{t+1} \rangle \}^{n-1}_{t=0}$
with $i_0 \in A, i_n \in B$ vanish. Indeed, if there is no such path, consider
the set of sites ${\cal X}(A)$ connected to $A$ by a path
in ${\cal X}$. By assumption, $B
\cap
{\cal X}(A)=\emptyset$. Now exchange $s^1_i$ and $s^2_i$ for all
$i \in {\cal X}(A)$. The
expectation  $\langle \cdot \rangle^0$ and each $f_{ij}$, for
$\langle ij \rangle \in {\cal X}$,
are even  under such an exchange, while $\tilde F$ is odd and $\tilde G$ is
even since  $B \cap {\cal X}(A)=\emptyset$.
 Hence, the corresponding term vanishes.

So we can choose a path $P$ connecting $A$ and $B$ for each non-zero term in
the numerator of (\ref{14}). We bound this numerator  as follows:
\be
\sum_P \sum_{{\cal X}}{^P}  \langle f_P \tilde F \tilde G
f_{{\cal X} \backslash P} \rangle^0
\leq \| \tilde F \| \; \| \tilde G \| \sum_P (4\beta J )^{|P|}
 \sum_{{\cal X}}{^P}
\langle f_{{\cal X} \backslash P} \prod_{\langle ij\rangle \in P}(1+f_{ij})
\rangle^0
\label{15}
\en
where we use (\ref{13}).
Observe that, by  the positivity of $f_{ij}$,
 the sum $\sum^P_{\cal X}$ is
bounded  by the sum over sets of $n.n.$ pairs in $\Lambda$ not belonging to
$P$  and so,
$$
\sum_{{\cal X}}{^P}  f_{{\cal X} \backslash P} \leq \prod_{
\langle ij \rangle \subset \Lambda , \langle ij \rangle \not \in P }
(1+f_{ij}).
$$
Therefore,
\be
\sum_{{\cal X}}{^P}
\langle f_{{\cal X} \backslash P} \prod_{\langle ij\rangle \in P}(1+f_{ij})
\rangle^0 \leq \langle \prod_{\langle ij \rangle \subset \Lambda} (1+f_{ij})
\rangle^0.
\label{16}
\en
Since the RHS of (\ref{16}) is the denominator of (\ref{14}) and since
all $P$'s in (\ref{15})
satisfy $|P| \geq d(A,B)$, we can use an estimate like
(\ref{9}) to prove (\ref{10}) for $\beta$ small.

A similar argument also shows that there is a unique Gibbs
state (i.e. a unique
phase). To prove that,
one introduces boundary conditions. Define
\be
{\cal H}(s_\Lambda|\hat s_{\Lambda^c}) = - J \sum_{\langle ij
 \rangle \subset \Lambda} s_is_j -
J \sum_{
\langle ij \rangle , i \in \Lambda , j \not \in \Lambda  } s_i
\hat s_j
\label{17}
\en
for any fixed configuration $\hat s_{\Lambda^c}$
outside $\Lambda$, and introduce the
corresponding expectation $\langle \cdot \rangle_{\Lambda, \hat s}$.
Uniqueness of
the Gibbs state means that for any two boundary conditions $s,s'$ and any
function $F$ as in (\ref{10}),
$\langle F \rangle_{\Lambda, s} - \langle F \rangle_{\Lambda,s'}
\to  0$ as $\Lambda \uparrow {\bf Z}^d$.

Introducing duplicate variables, we write
\ba
&&\langle F \rangle_{\Lambda,s} - \langle F \rangle_{\Lambda,s'}
=Z(\Lambda|s_{\Lambda^c})^{-1}Z(\Lambda|s'_{\Lambda^c})^{-1}\times\nonumber \\
&& \sum_{(s^1_i,s^2_i)i\in \Lambda} (F(s^1_A)-F(s^2_A)) \exp
(-\beta({\cal H}(s^1_\Lambda|s_{\Lambda^c}) +
{\cal H}(s^2_\Lambda|s'_{\Lambda^c}))).
\label{18}
\ea
Now, if we perform the same expansion as in (\ref{14}),
we see that any non-zero
term must contain a path connecting $A$ and the boundary of $\Lambda$. Hence,
repeating the arguments that led to (\ref{15}, \ref{16}), we get that
(\ref{18}) is bounded in absolute value by $C
\exp (-m d (A,\Lambda^c))$ for some $m>0$.

It is easy to extend this high-temperature expansion to two-body long range
interactions, with Hamiltonian $H=-\sum_{i,j} J_{ij} s_is_j$ and
\be
\sum_j |J_{ij}|e^{\gamma|i-j|} < \infty,
\label{19}
\en
 for some $\gamma > 0$. Now, the analogue of the connected paths entering
(\ref{15}) are sets of pairs $P = \{ ( i_t , j_{t} ) \}^{n}_{t=0}$
with $i_0 \in A$, $j_n \in B$ and $j_{t}=i_{t+1}$.
More generally, for interactions that have only
power-law decay, i.e. such that
$\sum_y |J_{ij}| |i-j|^\alpha < \infty$ some $\alpha > 0$, one gets
$\sum_k | \langle s_0s_k \rangle | |k|^\alpha < \infty$ for $\beta$ small.
The signs of $J_{ij}$ do not matter. We can always write
\ba
&&\exp (\beta J_{ij}( s^1_is^1_j+ s^2_is^2_j)) = \exp (-2\beta |J_{ij}|)
\exp (\beta (J_{ij}( s^1_is^1_j+ s^2_is^2_j)+2|J_{ij}|))\nonumber \\
&&=\exp (-2\beta |J_{ij}|) (1+f_{ij})
\label{20}
\ea
instead of (\ref{12}), so that the factors $f_{ij}$ are all positive, and,
using (\ref{19}), are bounded by $C\beta e^{-\gamma|i-j|}$ .

However, this does not work for {\em complex} interactions.
Actually, if we want
to prove analyticity (e.g. in $\beta$) of the
thermodynamic functions, we have to use
polymer or cluster expansions \cite{B,S}. While standard,
these methods are more involved
combinatorically than the arguments given above.

This dichotomy between expansions that prove exponential clustering for real
interactions and methods that imply analyticity becomes more relevant
 when we turn
to many-body interactions (involving arbitrarily large number of spins). These
interactions appear quite naturally, for example as effective
interactions after
performing a renormalization group transformation, or in
 dynamical systems, as we
shall see in the next sections. To discuss these interactions,
we need some definitions.
At each site of the lattice, we have a spin
$s_i \in \{0,\cdots,k-1\}$ (or, more generally, $s_i$ belongs to
a compact metric space). An interaction is defined
by a family $\Phi = (\Phi_X)$ of (continuous) functions of
the spin variables in $X$
indexed by finite
subsets $X$ of ${\bf Z}^d$.
We let $\|\Phi_X\|$ denote the sup norm of $\Phi_X$.
Given $\Lambda \subset {\bf Z}^d$, $|\Lambda| < \infty$, and a
configuration $ s'_{\Lambda^c} = (s'_i)_{i\in {\Lambda^c} }$ in
 $\Omega_{\Lambda^c}$, the
Hamiltonian in $\Lambda$ (with boundary conditions
 $ s'_{\Lambda^c}$) is defined as
\be
{\cal H}( s_\Lambda| s'_{\Lambda^c}) =
 -\sum_{X \cap \Lambda \neq \emptyset} \Phi_X ( s_{X
\cap \Lambda} \vee s'_ {X \cap \Lambda^c})
\label{21}
\en
where, for $X\cap Y= \emptyset$, $ s_{X } \vee s'_Y $
is the obvious configuration in
$X \cup Y$. The associated (finite volume) Gibbs measure is the probability
distribution :
\be \nu ( s_\Lambda | s'_{\Lambda^c}) = Z^{-1} (\Lambda|
s'_{\Lambda^c}) \exp (-{\cal H} ( s_\Lambda| s'_{\Lambda^c}))
\label{22}
\en
(we put a minus sign in
(\ref{21}) for convenience). See e.g. \cite{Ru,S,VFS} for more details on the
theory of Gibbs states.

Once one introduces many-body interactions,
 one has to distinguish carefully between different
 convergence conditions (norms)
replacing (\ref{19}) for two-body interactions. Since we have not put $\beta$
explicitely in (\ref{22}), high temperatures (= small $\beta$) will mean small
norm.
Here are some of the most common norms and the results that can be proven about
the corresponding interactions.
 \begin{enumerate}
\item[1)] If
\be
\| \Phi \|_1 = \sum_{0 \in X}
|X| \| \Phi_X \|
\label{23}
\en is small enough, then the Gibbs state is unique \cite{Do1,Do2,S}
\item[2)] If, for some $\gamma > 0$,
\be
\| \Phi \|_2 = \sum_{0\in X}
e^{\gamma d(X)}
\|\Phi_X\|,
\label{24}
\en
where $d(X)$ is the diameter of $X$, is small enough, then
the Gibbs state is unique and its
 correlation
functions decay exponentially \cite{Gr1}.
\item[3)] If, for some
$\gamma >0$,
\be
\| \Phi
\|_3 = \sum_{0 \in X} e^{\gamma|X|} \| \Phi_X \|
\label{25}
\en
is small enough,
then the Gibbs state
is unique and its correlation functions are analytic \cite{Is}.
\end{enumerate}

Note that for two-body interactions, (\ref{24})
reduces to (\ref{19}), and requiring the other norms to be
small is a much weaker condition in
that case. But, in general, as wediscussed in \cite{BK1}, $\|\Phi\|_2$ small
is not sufficient  to prove analyticity  of the thermodynamic functions. For
that, one basically needs $\|\Phi\|_3$ small. To see that  $\|\Phi\|_3$ small
does not necessarily imply  $\|\Phi\|_2$ small (or even finite), at least on
lattices
of more than one dimension, observe that for $X$ being a large square or a
large cube, $|X|$ is much larger than $d(X)$.
On the other hand, it is rather
easy to extend the arguments given above to prove exponential decay of the
correlation functions for a system with $\| \Phi \|_2$ small
enough (see Theorem 1 below).

For the applications to dynamical systems, one needs more than results for
 interactions with a small norm. High temperature expansion
means that one perturbs around a
completely decoupled system (i.e. the one defined by the expectation
$\langle \cdot \rangle^0$ in ({\ref 14})). However,
it is well-known  that one-dimensional systems
(say, with finite range interactions, i.e. with $\Phi_X =0$ if
$d(X)>R$ for some finite $R$) do not
undergo phase transitions, i.e. they remain in the
 high-temperature regime for all
temperatures. This results follows from the
 Perron-Frobenius theorem applied to the transfer
matrix of the system (see Proposition 1 below).
One might expect that this property is stable under small perturbations.
For
reasons which will be explained  in the next section, we shall need to
consider interactions which are the sum of two terms: a
one dimensional finite range interaction, whose norm is not necessarily
small,
and a more general
 interaction, which may be of infinite range, but whose $\| \Phi
\|_2$ norm is small.

Concretely, we will be interested
in the following setup.
Consider a lattice ${\bf Z}^{d+1}$ where we want to single out a particular
(``time") direction. Let us write $\alpha = (t,i)$ for
$\alpha \in {\bf Z}^{d+1}$,
$t\in {\bf Z}$ and $i \in {\bf Z}^d$. At each site $\alpha$, we have a spin
$s_\alpha \in \{0,\cdots,k-1\}$. The phase
space is a product of topological Markov chains (or subshifts of finite type):
$$
\Omega_{\cal A} = \{{\bf s} = (s_\alpha)_{\alpha \in {\bf Z}^{d+1}} |
{\cal A}(s_{(t,i)},s_{(t+1,i)}) = 1 \; \forall t \in {\bf Z}, \; \forall i \in
{\bf Z}^d\}
$$
and ${\cal A}(s,s')$ is a $k$ by $k$ matrix whose entries take value 0 or 1,
and which
is transitive  i.e., $\exists n < \infty$ such that ${\cal A}^{n} (s,s') >
0$  $\forall s,s' \in \{0,\cdots,k-1\}$. With this phase space, the expectation
values are defined as before, except that in (\ref{22}), we restrict $\nu$
to $ s_\Lambda$'s such that $ s_\Lambda \vee  s'_{\Lambda^c}$ belongs to
$\Omega_{\cal A}$, and, in $Z^{-1} (\Lambda|
s'_{\Lambda^c})$, the sum runs over those $ s_\Lambda$'s. From the point of
view of statistical mechanics, ${\cal A}$ defines a nearest-neighbor
hard-core along the time direction, where the forbidden configurations
are those where
a pair satisfies
${\cal A}(s_{(t,i)},s_{(t+1,i)}) = 0 $.

Let $\Phi^0$ be a finite range one-dimensional interaction,
i.e. $\exists R < \infty$
such that $\Phi^0_X=0$ if diam $(X) > R$ and $\Phi^0_X=0$
 unless $X \subset {\bf Z}
\times \{i\}$ for some $i \in {\bf Z}^d$, i.e. unless $X$
is a subset of a one-dimensional sublattice of ${\bf Z}^{d+1}$ along the
``time" direction.
 Without loss of generality, we may assume that $R>n$.

\begin{Th} Let $\Phi = \Phi^0 + \Phi^1$, where $\Phi^0$ is as above. Then,
there exist $\epsilon >0$, $m>0$, $C< \infty$ such that, if $\| \Phi^1 \|_2
\leq \epsilon$,
there is a unique Gibbs state $\mu$ for $\Phi$ and the correlation functions
satisfy, for
all $F$, $G$, with $F : \Omega_A \to {\bf R}$, $G : \Omega_B \to {\bf R}$,
where $A,B
\subset {\bf Z}^{d+1}$ are finite:
\be | \langle FG \rangle - \langle F \rangle \langle G
\rangle |  \leq C \min (|A|,|B|) \|F\|\|G\| e^{-md({A},{B})},
\label{26}
\en
where $d({A},{B})$ is the
distance between the sets $A$ and $B$ and $\langle F \rangle =\int F d\mu$.
\end{Th}

\vspace*{4mm}

\par\noindent {\bf Remark 1.} We refer to \cite{BK1} for an
alternative version
 of this Theorem, and for various extensions. In \cite{BK1},
we took $\Phi^0$ to be any interaction which is in a
suitably defined high-temperature regime. Note, however,
 that in \cite{BK1}, we had
${\cal A}(s,s')=1$ for all $s,s'$.

\vspace*{4mm}

\par\noindent {\bf Remark 2.}
Since $\Phi^0$ has no coupling in spatial directions
 and has range $R$ in the time
direction we get the following decomposition of the corresponding
 partition
function $Z^0$.  Given $V \subset {\bf Z}^{d+1}$, we let, for
$i\in {\bf Z}^d$, $J_i= ({\bf Z}\times i) \cap V$ so that
$V = \cup_i J_i$. We shall call the $J_i$'s the time intervals of $V$.
 Then,
 \be Z^0 (V |
{s}) = \prod_{i} Z^0 (J_i | s).
\label{27}\en

The transfer matrix formalism gives  the following representation
for $Z^0 (J | s)$:
\begin{Pro} There exist $\gamma > 0$, $C < \infty$ independent
of $R$ such that, if $J $
is an interval of the form $[(i,t),(i,t+ \ell R)]$, $\ell \in {\bf N}$,
 then \be
 Z^0(J|s) = \lambda^\ell W ({
s}_-) W(s_+) (1+ g_J (s_-,s_+))
\label{28}
\en
where
$s_+=s_{((i,t+\ell R),(i,t+ (\ell +1)
R)]}$ and $s_-=s_{[(i,t- R),(i,t))}$ with \be | g_J (s_- , s_+) | \leq C
e^{-\frac{\gamma}{R}|J|}.
\label{29}
\en
\end{Pro}
This last Proposition is rather
standard in statistical mechanics; for a proof, see for example
the Appendix of
 \cite{BK1}, where one has to replace ${\cal T}$, $P$ by
${\cal T}^n$, $P^n$, to take into account the presence of the matrix
${\cal A}$.
\vspace*{4mm}

\par\noindent {\bf Remark 3.} We could write $s_{\pm}(J)$ to indicate
that the configurations $s_{\pm}$ are indexed by sites situated on both sides
of the interval $J$. Keeping that in mind, we write $s_{\pm}$, for simplicity.

\vspace*{4mm}

\vspace*{3mm} \par\noindent {\bf Proof of Theorem 1.} We shall sketch the proof
of
 (\ref{26}), using the ideas explained in this section,
and refer the reader to \cite{BK1}
for more details.
First, we write the LHS (the
expectation are taken with respect to a finite volume Gibbs state with open
boundary
conditions) of (\ref{26})  using duplicate variables:
\ba
&&2(\langle FG
\rangle_\Lambda - \langle F \rangle_\Lambda \langle G \rangle_\Lambda)\nonumber
\\
&=& Z(\Lambda)^{-2}
\sum_{ s^{i}_\Lambda,i=1,2} \tilde F \tilde G
\exp (-{\cal H}( s^1_\Lambda) - {\cal H}(
s^2_\Lambda))
\label{30}
\ea
where $\tilde F = F ( s^{1}_\Lambda) - F( s^{2}_\Lambda),
\tilde G = G(
s^{1}_\Lambda) - G ( s^{2}_\Lambda)$.

We may replace $\Phi^1_X$ by $\Phi^1_X - \inf_{ s_X} \Phi^1_X$, by
adding a constant to
the Hamiltonian. Thus, we may, without loss of generality,
assume that $\Phi^1_X \geq 0$
for all $X$, and that $ \Phi^1$ still satisfies
$\| \Phi^1 \|_2 \leq 2 \epsilon$. Then, we
perform a  high-temperature expansion on the $\Phi^1$
 part of ${\cal H}$, as in (\ref{14}):
\be
\exp(\sum_{X,i=1,2} \Phi^1_X ( s^{i}_X)) = \sum_{\cal X} \prod_{X \in {\cal X}}
f_X
\label{31}
\en
where the sum runs over sets ${\cal X}$ of subsets of $\Lambda$, and
\be
f_X = \exp (\Phi^1_X (s^1_X) + \Phi^1_X ( s^2_X)) - 1
\label{32}
\en
satisfies:
\be
0 \leq f_X,
\label{33}
\en
\be
\sum_{0\in X}e^{\gamma d(X)} \| f_X \| \leq C \epsilon
\label{34}
\en
(using the positivity of $\Phi^1_X$, (\ref{24}) and $\| \Phi^1 \|_2
\leq\epsilon$).

It is convenient to cover ${\bf Z}^{d+1}$
by disjoint $R$-intervals, i.e. intervals of length $R$
parallel to the time axis.
Now, insert (\ref{31}) in (\ref{30}) and, for each term in (\ref{31}),
define $V =V({\cal X}) =
\underline{A} \cup \underline{B} \cup \underline{{\cal X}}$,
where $\underline{\cal
X}=\cup_{X \in {\cal X} } \underline{X} $ and for any $X\subset {\bf
Z}^{d+1}$, $\underline{X}$ is the set of $R$-intervals intersected by $X$.
We have, by summing, in
(\ref{30}), first over $s^{i}_{\alpha}$, for
 ${\alpha}\in \Lambda \backslash V$, and then over
$s^{i}_{\alpha}$,  ${\alpha}\in V$,
\be
(\ref{30})=
Z(\Lambda)^{-2} \sum_{\cal X} \sum_{ s^{i}_V,i=1,2} \tilde F
\tilde G \prod_{X \in {\cal
X}} f_X \exp (-{\cal H}^0_V) \prod_{i=1,2} Z^0
(\Lambda \backslash V | s^{i}_V) \label{35}\en
where
${\cal H}^0_V=-\sum_{X \subset V} (\Phi^0_X ( s^1_X) + \Phi^0_X
( s^2_X))$, and $ Z^0
(\Lambda \backslash V | s^{i}_V)$ is the partition function with
interaction $\Phi^0$, $
s^{i}_V$ boundary condition in $V$, and open boundary conditions
in $\Lambda^c$.
Assuming $\Lambda$ to be a union of $R$-intervals (it is easy to extend the
proof to the general case), $\Lambda \backslash V$ is also such a union, and
we may use (\ref{27},\ref{28}) for $Z^0 (\Lambda \backslash V | s^{i}_V)$.
Since $Z^0 (J|s)$ does not vanish for $J$ being a union of $R$-intervals (we
chose $R>n$), we may define
$  1+ {\tilde g}_J (s^1_- , s^1_+, s^2_- , s^2_+)
= \prod_{i=1,2}\frac{1 + g_J (s^{i}_-, s^{i}_+)}
{\min_{s^{i}_-, s^{i}_+}(
1 + g_J (s^{i}_-, s^{i}_+))} $ so that ${\tilde g}_J (s^{i}_- , s^{i}_+)$ is
positive and still satisfies
(\ref{29}) (with a different $C$):
\be
0\leq {\tilde g}_J (s^{i}_- , s^{i}_+) \leq C
e^{-\frac{\gamma}{R}|J|}.
\label{36}
\en

We get:
$$
Z^0
(\Lambda \backslash V | s^{i}_V)={\cal
W}(s^{i}_V)\prod_{J}(1 + \tilde g_J (s^{i}_-, s^{i}_+))
$$
where
\be
{\cal
W}(s^{i}_V)=\prod_{J}\lambda^{2|J|} \prod_{i=1,2} W ( s^i_{-})
W ( s^i_{+})
\min_{s^{i}_-, s^{i}_+}(
1 + g_J (s^{i}_-, s^{i}_+)).
\label{38}
\en
The product over $J$ runs over all the time intervals of $\Lambda
\backslash V$ (so that the variables $s^1_{\pm}$, $s^2_{\pm}$ are indexed by
sites in $V$).

Let us insert this representation of  $Z^0
(\Lambda \backslash V | s^{i}_V)$
in (\ref{35}) and then
expand the product over $J$
of $(1 + \tilde g_J (s^{i}_-, s^{i}_+))$. The result is:
 \be
(\ref{35})= Z(\Lambda)^{-2} \sum_{{\cal X},{\cal J}}
\sum_{ s^{i}_V, i=1,2} \tilde F \tilde G \prod_{X \in {\cal X}}
f_X \prod_{J \in {\cal J}}
\tilde g_J \exp (-{\cal H}^0_V) {\cal W}(s^{i}_V)
\label{37}
\en
where  the sum over
${\cal J}$  runs over families
of intervals
$ J \subset \Lambda
\backslash V$.

{}From now on,  we can proceed as we did previously:
the main observation is again that
 if, for each term in (\ref{37}),
 we decompose $V\cup (\cup_{\cal J}J)$ into connected components (where
connected is defined in an obvious way: any two sets can be joined by a
``connected path" $P =
(Z_i)^n_{i=1}$ where each $Z_i$ is either an ${\underline X} $ or a $J$ and
the distance between $Z_{i+1}$  and  $ Z_i$ is less than $1$, $\forall i =
1,\cdots,n-1$),
 and
if $A$ and $B$ belong to different components, then that term vanishes.
Let us check this: as before, we interchange $s^1_{\alpha}$ and $s^2_{\alpha}$
for each
${\alpha}$ in the connected component containing $A$.
 $\tilde F$ is odd under such
an
interchange, while $\tilde G$ is even (if $A$ and $B$ belong to different
components), and $f_X$, $\tilde g_J$ are obviously even. Next, observe that
${\cal
W}(s^{i}_V)$ can be factorized into a product of functions, each of which
depends only of $s^i _{\alpha}$ for ${\alpha}$ belonging to a
connected component of $V$. Observe also that
${\cal H}^0_V$ does not contain terms where
$X$ intersects different connected components of $V$ (since $\Phi_0$ has range
$R$ and $V$ is defined as a union of $R$-intervals). So, the two last factors
in (\ref{37}) factorize over connected components of $V$, and are
therefore also even under our interchange of $s^1_{\alpha}$ and $s^2_{\alpha}$.

Hence, for each non-zero term in (\ref{37}), we can
choose a connected path, as defined above, $P =
(Z_i)^n_{i=1}$ where $Z_1 = {\underline A}, Z_n =
{\underline B}$. Then, using
the positivity of $f_X$, $\tilde g_J$, we bound the sum
in  (\ref{37}) by
a sum over such paths, and control that sum essentially as in (\ref{9}).
The exponential decay comes from combining (\ref{34}), when $Z_i$ is
an ${\underline X} $
and
(\ref{36}) when $Z_i$ is
a $J $.
 The uniqueness of the Gibbs
state is proven as in (\ref{18}); for details, see \cite{BK1}.\hfill$
\makebox[0mm]{\raisebox{0.5mm}[0mm][0mm]{\hspace*{5.6mm}$\sqcap$}}$ $ \sqcup$
\vspace*{3mm}
\vspace*{3mm}

\setcounter{equation}{0}
\section{SRB measures for expanding circle maps.}
\vspace{5mm}

We start by recalling the standard theory of invariant measures for smooth
expanding circle maps, in a formulation that will be
used later. To describe the dynamics, we first fix a map $F: S^1 \to S^1$. We
take  $F$ to be an
expanding, orientation preserving $C^{1+\delta}$ map with $\delta >0$
 (i.e. $F$ is differentiable and its derivative is H\"older continuous of
exponent
$\delta$).
 We describe $F$ in
terms of its lift to ${\bf R}$, denoted by $f$ and chosen,
say, with $f(0) \in [0,1[$.  We
assume that
\be f'(x) > \lambda^{-1}
\label{2.1}
\en
 where $\lambda < 1$. Note that
 there exists an
integer $k > 1$ such that
\be f(x+1) = f(x) + k \;\;\;\; \forall x \in
{\bf R}.
\label{2.2}
\en

A probability measure $\mu$ on $S^1 $ is called an SRB measure if it is
$F$-invariant and absolutely continuous with respect to the Lebesgue
measure.
The following results are well-known for maps $F$ as above
(see \cite{Si2,Co,KH}):
\begin{enumerate}
\item[(a)] There is a unique SRB measure $\mu$.
\item[(b)] For any absolutely continuous probability measure $\nu$, and any
continuous
function $G$,
\be
\int G \circ F^N d\nu \rightarrow \int G d\mu
\label{2.3}
\en
as $N \rightarrow \infty$.
\item[(c)] There exists $C<\infty$, $m>0$, such that
$\forall G \in L^{\infty}(S^1), \forall H \in {\cal C}^\delta (S^1)$,
\be |\int G \circ F^n Hd\mu - \int G d \mu \int H d \mu| \leq C \| G
\|_\infty \| H \|_\delta e^{-mn},
\label{2.4}\en
where ${\cal C}^\delta (S^1)$ denotes the space of H\"older continuous
functions, with the norm
$$
\| H \|_\delta = \| H \|_\infty + \sup_{{x,y}} \frac{|H({x}) - H
({ y})|}{ |x - y |^\delta}.
$$
\end{enumerate}

\vs{3mm}

\par\noindent {\bf Remark.} There are different ways to define an SRB measure.
In \cite{ER}, they are introduced as  measures whose restriction on the
expanding
directions is absolutely continuous with respect to the Lebesgue
measure. Since here the whole phase space $S^1$ is expanding, our definition is
natural
(besides, with this definition, the SRB measure is unique). But, as
we mentioned in the introduction, one of the most interesting properties of the
SRB measure
is that it describes the statistics of the orbits of almost every point, which
means that
\be
\frac{1}{N}\sum_{i=0}^{N-1} G \circ F^i (x) \rightarrow \int G d\mu,
\label{2.5}
\en
for almost every $x$, and every continuous function $G$. Note that, if we
integrate (\ref{2.5}) with an
absolutely continuous probability measure
$\nu$, we obtain the Cesaro average of (\ref{2.3}). So, both properties are
related to each other.

\vspace*{3mm}

Let us sketch now the construction of $\mu$. In doing so, we shall establish
the connection with the statistical mechanics of one-dimensional spin
systems. This way of constructing $\mu$
may not be the simplest one in the present context,
but the
connection to statistical mechanics will be essential in the
analysis of coupled maps (see \cite{Si2,Co,KH}
for different approaches, although the one below is close to \cite{Si2}).

The Perron-Frobenius operator $P$ for $F$ is defined  by \be
\int G \circ F H d m = \int G P H dm
\label{2.6}
\en for
$G \in L^\infty (S^1)$, $ H \in L^1 (S^1)$, and $dm$ being the Lebesgue
measure. Let us work in the covering space ${\bf R}$ and replace $G,
H$ by periodic functions denoted $g, h: g ( x + n) = g ( x)$, $
\forall n\in {\bf Z}$. We get
 \be
\int_{[0,1 ]} g \circ f h d {x} = \int_{[0,1 ]} g (
 x) P h ({ x}) d{ x}.
\label{2.7}
\en

More explicitely,
\be
 P h ({
x}) = \sum_{{ s}} \frac{h(f^{-1} ({ x} + { s}))}{ f' (f^{-1} (x +
s))}
\label{2.8}
\en
 where ${ s} \in \{0,\cdots,k-1\}$ (and $k$ was introduced
in (2)). Note that $P$ maps periodic functions into periodic
functions because the sum is periodic even if the summands are not: indeed,
(2) implies that  $f'$ is periodic and that
$f^{-1} (x+1+k-1) = f^{-1} (x+k) =
f^{-1} (x) + 1$ (so that, if we add $1$ to $x$, it amounts to a cyclic
permutation of $s$).

By (\ref{2.7}), the density $h_{\mu}(x)$ of the absolutely
 continuous invariant measure $d\mu =h_{\mu}(x)
dx$ satisfies $Ph_{\mu}=h_{\mu}$. We shall construct
$h_{\mu}$ as the limit, as $N \rightarrow
\infty$, of $P^N 1$.  $P^N 1$ has a direct statistical mechanical
interpretation which we now derive.

First, iterating (\ref{2.8}), we get
\be (P^N 1) ({ x}) = \sum_{{ s}_1, \cdots, {
s}_N} \prod^N_{t=1} [  f' (f^{-1}_{
s_t} \circ \cdots \circ f^{-1}_{ s_1} ({ x}))]^{-1}
\label{2.9}
\en
where $ f^{-1}_{s} (x) \equiv f^{-1}(x+s)$.

{}From now on,  we shall
consider ${ x}\in [0,1 ]$.
We introduce a convenient notation: ${ x}\in [0,1 ]$  and ${ s}_1,
\cdots, { s}_N$ in (\ref{2.9}) collectively define a configuration on a
 lattice $\{0,\cdots,N\}$.  To any subset $X \subset {\bf
Z}_+$ associate the configuration space $\Omega_X = \times_{t \in
X} \Omega_t$ where $\Omega_t$ equals [0,1] if
$t=0$, and equals $\{0,\cdots,k-1\}$ if $t>0$. We could
use the existence of a
Markov partition for $F$ to write ${ x}$ as a symbol sequence, as
is usually done, e.g. in \cite{BS}, but we shall not use explicitely this
representation.

Let
 ${\bf s} = ({ x},{ s}_1,
\cdots,{ s}_N) \in \Omega_{N}$.
 Then (\ref{2.9}) reads
\be (P^N 1) ({ x}) = \sum_{{ s}_1 \cdots { s}_N} e^{- {\cal
H}_N({\bf s})}
\label{2.10}
\en
 with $e^{- {\cal
H}_N({\bf s})} $ being the summand
in (\ref{2.9}). And we want to construct the limit:
\be
\int_{[0,1 ]} g(x) d\mu=
\lim_{N \rightarrow
\infty}\int_{[0,1 ]} g(x)(P^N 1) ({ x})dx = \lim_{N \rightarrow
\infty}\sum_{{ s}_1 \cdots { s}_N}\int_{[0,1 ]}g(x) e^{- {\cal
H}_N({\bf s})}dx
\label{2.11}
\en
for any continuous function $g$.

This is the statistical mechanical representation we want to use.
In that language, $d\mu$ is the restriction to  the ``time zero"
phase space of
the Gibbs state
determined by ${\cal
H}$.
One can also rewrite the time correlation functions (\ref{2.3}) as follows:
let $d \nu =h_\nu (x) dx$; then, replacing again $G$ by a periodic
function $g$ and $F$ by its lift,
\be
\int_{[0,1 ]} g \circ f^N h_\nu dx = \int_{[0,1 ]} g P^N
h_\nu d x  = \sum_{({ s}_i)_{i=1}^N} \int_{[0,1 ]} g
( x) \exp (- {\cal H}_{N} ({\bf s})) h_\nu ({\bf s}) dx
\label{2.12}
\en
where
$$
 h_\nu ({\bf s})= h_\nu (f^{-1}_{
s_N} \circ \cdots \circ f^{-1}_{ s_1} ({ x})),
$$
and the last equality in (\ref{2.12}) follows by iterating (\ref{2.8}).

On the other hand, since $\int Ph dx =\int h dx$, by definition of $P$
(use (\ref{2.7}) with $g=1$),
one has
\be
\sum_{({ s}_i)_{i=1}^N} \int_{[0,1 ]}
 \exp (- {\cal H}_{N} ({\bf s})) h_\nu ({\bf s}) dx = \int_{[0,1 ]}  P^N
h_\nu d x=\int_{[0,1 ]} h_\nu dx =1,
\label{2.13}
\en
 since $d\nu$ is a probability measure. So, using (\ref{2.11}, \ref{2.12}),
one sees that (\ref{2.3}) is translated, in the statistical mechanics language,
 into
\ba
&&\lim_{N \rightarrow
\infty}(\int G \circ F^N d\nu - \int G d\mu)
\nonumber\\
&=&\lim_{N \rightarrow\infty}(\int_{[0,1 ]} g P^N
h_\nu d x-\int_{[0,1 ]} g P^N
1 d x )
\nonumber\\
&=&\lim_{N \rightarrow
\infty}(\sum_{({ s}_i)_{i=1}^N} \int_{[0,1 ]} g
( x) e^{- {\cal H}_{N} ({\bf s})} h_\nu ({\bf s}) dx\nonumber\\
&-&
(\sum_{({ s}_i)_{i=1}^N} \int_{[0,1 ]} g
( x) e^{- {\cal H}_{N} ({\bf s})} dx)
(\sum_{({ s}_i)_{i=1}^N} \int_{[0,1 ]}
e^{- {\cal H}_{N} ({\bf s})} h_\nu ({\bf s}) dx))
\nonumber\\
&=& 0
\label{2.14}
\ea
where we used (\ref{2.13}) to insert the last factor (which equals one). We
shall see below that the last equality
 expresses the  decay of correlation functions
for the Gibbs state determined by ${\cal H}$. A similar observation
 holds for (\ref{2.4}).

One advantage of these representations is that one may use
the statistical mechanics formalism to control the limit.
 To make the connection with statistical mechanics more explicit,
 it is convenient to write ${\cal H}_N$ in terms of
many-body interactions.
First of all, from (\ref{2.9}, \ref{2.10}), we get
\be
{\cal H}_N=\sum_{t=1}^N V_t
\label{2.15}
\en
with
\be
V_t ({\bf s})=  \log (f' (f^{-1}_{
s_t} \circ \cdots \circ f^{-1}_{ s_1} ({ x}))).
\label{2.151}
\en
We may localize $V_t$ by writing it as a telescopic sum:
\be
V_t ({\bf s})= \sum_{l = 0}^{t}
\Phi_{[l,t ]} ({\bf s}) + V_{t} ( {\bf 0}_{[0,t ]})
\label{2.16}
\en
where, for $l\neq 0,t$,
\be \Phi_{[l,t ]} ({\bf s}) = V_{t} ({\bf s} _{[l,t ]}
\vee { \bf 0}_{[0,l-1 ]}) -  V_{t} ({ \bf s} _{[l+1,t ]}
\vee {\bf 0}_{[0,l ]})
\label{2.17}
\en
and ${\bf 0}_{[0,l ]}$ denotes the configuration equal to $0$
for all $i\in [0,l ]$
(note that $0$ belongs to the
phase space for all $i$'s). For $l=0,t$, $\Phi_{[l,t ]} ({\bf s})$ is
 given by a similar formula, where the intervals that would appear in
(\ref{2.17}) as $[0,-1 ]$
and $[t+1,t ]$ are replaced
 by the empty set.

Combining (\ref{2.15}, \ref{2.16}), we may write
the Hamiltonian as a sum of many-body interactions:
\be
{\cal H}_N({\bf s})=\sum_{t=1}^N \sum_{l = 0}^{t}
\Phi_{[l,t ]} ({\bf s}) + C
\label{2.18}
\en
where the constant $C=\sum_{t=1}^N V_{t} ( {\bf 0}_{[0,t ]})$.

The main point of (\ref{2.18}) is that $\Phi_{[l,t ]}
({\bf s})$ depends on $\bf s$
only through ${\bf s}_{[l,t ]}$. In the statistical
 mechanics language, these are
many-body interactions  coupling all
the variables in the interval $[l,t ]$. The next Proposition
 shows that these interactions decay
exponentially with the size of the interval $[l,t ]$.

\begin{Pro} There exists $C< \infty$, such that
\be |
\Phi_{[l,t ]} ({\bf s}) | \leq C \lambda^{\delta (t-l)}.
\label{2.19}
\en
\end{Pro}
{\bf Proof.} This combines two bounds:
First, since $F$ is  $C^{1+\delta}$, one has
\be
|\log f' (x) -\log f' (y)|\leq C |x-y|^\delta
\label{2.191}
\en
and, by (1),
\be
|f^{-1}_s (x) -f^{-1}_s (y)|\leq \lambda |x-y|.
\label{2.20}
\en
Then, iterating (\ref{2.20}), one gets:
$$
|f^{-1}_{
s_t} \circ \cdots \circ f^{-1}_{ s_{l+1}} ({ x})-f^{-1}_{
s_t} \circ \cdots \circ f^{-1}_{ s_{l+1}} ({ y})|
\leq  \lambda^{t-l}
$$
since $|x-y|\leq 1$.
Then (\ref{2.19}) follows from this and (\ref{2.191}), since the $s$
variables in both terms of
 (\ref{2.17}) (see (\ref{2.151})) coincide in the first
$t-l$
places.\hfill$
\makebox[0mm]{\raisebox{0.5mm}[0mm][0mm]{\hspace*{5.6mm}$\sqcap$}}$
$ \sqcup$
\vspace*{3mm}

We can formulate the system here in the language of Section 2 as follows:
we write the interaction as the sum of an interaction $\Phi^0$ of
finite range $R$, which does not necessarily have
a small norm plus a long range
``tail" $\Phi^1$ whose norm can be made as small
as we wish by choosing $R$ large
enough. Concretely,
choose now $R$  to be the smallest integer such that
\be
\lambda^{\frac{\delta R}{2}}<
\epsilon.
\label{2.21}\en
Then, we  define $\Phi^0$ as grouping all the
$\Phi_{[l, t]}$'s with $t-l \leq R$
and $\Phi^1$ to collect all the longer range $\Phi_{[l, t]}$'s.
Since, for an interval $X= [l, t]$, $d(X)= t-l$, we easily have (for all
$s\in {\bf Z}_+$) the bound:
\be
\sum_{X \ni s} e^{\gamma
d(X)} \| \Phi^1_{ X} \| \leq C \epsilon
\label{2.22}
\en
for $\gamma$ small enough (e.g. so that $e^{\gamma}
\leq\lambda^{\delta/2 }$) ,
and where $C$ depends on
 $\lambda^{\delta }$.
Note also that, here, $d(X)=|X|-1$, so that, in this one-dimensional
situation, the norms
(2.24) and (2.25) are equivalent.
However, we cannot use Theorem 1 directly, because the way this Theorem is
stated, $\epsilon$ depends on $\Phi^0$, i.e. on $R$, and, here, we choose
$R$ in (\ref{2.21}) in an $\epsilon$-dependent way. It turns out that all we
would need
in the proof of Theorem 1 is
that
$R^{d+1} \epsilon$ is small enough (with $d=0$ here), and that is compatible
with (\ref{2.21}),  for $R$ large (see \cite{BK1} for details). Of course, in
this example,  one can also apply directly the transfer matrix formalism to
infinite-range interactions decaying as in (\ref{2.19}), see \cite{Ru}.

In order to prove the decay of the correlation functions
(\ref{2.3},\ref{2.4}) one proceeds as follows.
First, note that we can approximate the $L^1$ function $h_\nu$,
in the $L^1$ norm,
 by a smoother function, ${\tilde h_\nu}$, e.g. by
a H\"older continuous function of exponent $\delta$. Since $G$ in
 (\ref{2.3},\ref{2.14}) is bounded, this
means that we have the following approximation, uniformly in $N$:
$$
|\int G \circ F^N h_\nu dx -\int G \circ F^N {\tilde h_\nu} dx|
\leq \|G \|_{\infty} \| h_\nu - {\tilde h_\nu}\|_1.
$$
So, it is enough
to prove (\ref{2.14}) when $h_\nu$   is H\"older
continuous.

Thus, we may write a telescopic sum, as in (\ref{2.16}):
\be
h_\nu({\bf s})= \sum_{l = 0}^{N}
h_{\nu,[l,N ]} ({\bf s}) + h_\nu ( {\bf 0}_{[0,N ]})
\label{2.23}
\en
and one has an exponential decay
of the form:
\be
|h_{\nu,[l,N ]} ({\bf s})|\leq C\lambda^{\delta (N-l)}
\label{2.231}
\en
 as in
Proposition 2, since $h_\nu$ is  H\"older continuous. Now insert
(\ref{2.23}) in (\ref{2.14}), observe that $g(x)$ depends only
on the ''time zero" variable $x$, while
$h_{\nu,[l,N ]} ({\bf s})$ depends only on ${\bf s}_{[l,N ]}$.
Since (\ref{2.14}) has the form of correlation function, we can use
the exponential decay of the Gibbs state determined by ${\cal H}$, i.e. (2.26)
with $A=0$ and $B=[l,N ]$, hence $d(A,B)= l$.
Combining this exponential decay
with the exponential decay of $h_{\nu,[l,N ]} ({\bf s})$ and with
\be
\sum_{l=0}^N e^{-ml}\lambda^{\delta (N-l)}\leq  Ce^{-m'N}
\label{2.24}
\en
for $m'<\min (m,\delta |\log \lambda|)$, one proves (\ref{2.3}).

If $h_\nu$ is not H\"older continuous, the limit in (\ref{2.3}) is still
reached (via our approximation argument), but not necessarily exponentially.
The proof of
(\ref{2.4}) is similar. One sees also why, in (\ref{2.4}), one requires $H$ to
be  H\"older continuous,
while $G$ is only bounded: in order to prove exponential decay, we had to use
(\ref{2.23}, \ref{2.231}).

\setcounter{equation}{0}
\section{Coupled map lattices.}

We consider now a lattice of coupled expanding circle maps.
 The phase space ${\cal M} =
(S^1)^{{\bf Z}^d}$ i.e. ${\cal M}$ is the set of maps
${\bf z} =(z_j)_{j\in {\bf Z}^d}$
from ${\bf Z}^d$ to the circle.

To describe the dynamics, we first consider a map $F: S^1 \to S^1$ as in
Section 3. We let
${\cal F}: {\cal M} \to {\cal M}$ denote the Cartesian
product ${\cal F} = {\rm X}_{i \in
{\bf Z}^{d}} F_i$ where $F_i$ is a copy of $F$.
${\cal F}$ is called the uncoupled map.

The second ingredient in the dynamics is given by the
coupling map $A: {\cal M} \to {\cal
M}$. This is taken to be a small perturbation of the identity in
the following sense. Let
$A_j$ be the projection of $A$ on the j$^{\mbox{th}}$ factor and
let $a_j$ denote the lift of $A_j: A_j = e^{2 \pi i a_j}$. We take, for
example,
 $$
a_j({\bf x}) = x_j + \epsilon \sum_k g_{|j-k|} (x_j , x_k) $$
where $g$ is a
periodic $ C^{1+\delta}$ function in both variables, with exponential
falloff in $|j-k|$. We shall come back later on the reasons for
considering this somewhat
unusual model (see Remark 3 below).
 More general examples of such $A's$ can be found in
\cite{BK1,BK2} (note, however, that in \cite{BK2},
we restricted ourselves to analytic maps).

\vs{3mm}

The coupled map $T: {\cal M} \to {\cal M}$ is now defined by
$$T = A \circ {\cal F}.$$ We
are looking for ``natural" $T$-invariant measures on
 ${\cal M}$. For this, write, for
$\Lambda \subset {\bf Z}^d$, ${\cal M}_\Lambda =
(S^1)^\Lambda$, and let $m_\Lambda$ be
the product of Lebesgue measures.

\vspace*{5mm} \par\noindent {\bf Definition 1} {\em
 A Borel probability measure $\mu$ on
${\cal B}$ is a SRB measure if
\begin{enumerate}
\item[(a)] $\mu$ is $T$-invariant
\item[(b)] The restriction $\mu_\Lambda$ of $\mu$ to ${\cal B}_\Lambda$
is absolutely
continuous with respect to  $m_\Lambda$ for all
$\Lambda \subset {\bf Z}^d$ finite.
\end{enumerate}}

\vspace*{4mm} \par\noindent {\bf Remark 1.} This is a natural
 extension to infinite dimensions of the notion of SRB
measure, given in section 3, since each $S^1$ factor can
be regarded as an expanding
direction.
However, unlike the situation for single maps, we do not show
that the SRB measure is unique (although we expect it to be so). In
\cite{BK1}, we prove a weaker result, namely
that there is a unique ``regular" SRB measure. We also show that (3.3)
holds for $\nu$ being a ``regular" measure, but we have not extended
(3.5). The extension of (3.4)
is given below.

\vspace{2mm}

Our main result is:
 \begin{Th} Let $F$ and $A$ satisfy the assumptions given above. Then there
exists $\epsilon_0 > 0$ such that, for $\epsilon < \epsilon_0$, $ T$ has an SRB
measure
$\mu$. Furthermore, $\mu$ is invariant and exponentially mixing under the
space-time
translations: there exists $m>0$,
$C<\infty$, such that, $\forall B,D \subset {\bf Z}^d, |B|,|D| < \infty$ and
$\forall G \in L^{\infty}({\cal M}_B), \forall H \in {\cal C}^\delta ({\cal
M}_D)$,
\be
|\int G \circ T^n Hd\mu - \int G d \mu \int H d \mu| \leq C \| G
\|_\infty \| H \|_\delta  e^{-m(n+d(B,D))},
\label{3.1}
\en
 where $d(B,D)$
is the distance between $B$ and $D$ and $C$ depends on $d(B), d(D)$.
\end{Th}

\vspace*{4mm}\no {\bf Remark 2.}
The proof combines the ingredients from the previous two sections.
We first derive a formula for the Perron-Frobenius operator of
$T$ which is similar to (3.9, 3.10). And we express the Hamiltonian
in terms of
potentials as in (3.19), using a telescopic sum. The decay of the potentials
is proven again using the H\"older continuity of the differential of $T$
and the
expansivity of $F$. We may write the potential $\Phi$ as a sum of two terms,
as in Theorem 1, $\Phi^0 + \Phi^1$, with $\Phi^0$ one-dimensional and of
finite range
 and $\| \Phi^1 \|_2 $ small, but
on a ``space-time" ${\bf Z}_+^{d+1}$ lattice. And,
 using Theorem 1 (which can trivially be extended to this lattice), we
construct the SRB measure and prove the exponential decay of correlations.
For a discussion of previous work on this problem, see \cite{BK1}.

\vspace*{4mm}\no {\bf Remark 3.}
 One would like to extend this Theorem to coupled maps of the interval
$[0,1]$ into itself, where the uncoupled map is not smooth, but, say,
of bounded
variation. Indeed, all examples were phase transitions are expected to occur
 are of this form (see e.g. \cite{MH,Po}). Moreover, the theory for a single
map can easily be extended to maps of bounded variation \cite{Co}.
Also, one would like to consider more general couplings $A$, like the standard
diffusive coupling.

However, such extensions seem rather difficult, because even if the uncoupled
map happens to have a Markov partition, the couplings tend to destroy these
partitions. This is basically the reason for
considering circle maps instead of expanding maps of the interval. We did
not use explicitely the existence of a Markov partition, but we used it
implicitely  because no characteristic functions appeared in the formula (3.9)
for
 the Perron-Frobenius operator (compare with  the formula for $P$ in
\cite{Co}).  The
reader should not be misled  by the fact that, in the statistical mechanics
part of the argument  (Section 2), we could handle a general transitive matrix
$\cal A$, defining a subshift. Indeed this is a short-range hard-core
interaction,  in the statistical
mechanics language, while the appearance of characteristic functions
in  the Perron-Frobenius operator
 may give rise to an infinite range hard
core, and this is much more difficult to control.

Note, however, that existence results on SRB measures in
this more general context
were obtained in \cite{K}. But there are no results on the
 exponential decay of correlation functions. Also, Blank has constructed
examples of ``pathological" behaviour for  coupled non-smooth maps with
arbitrarily weak coupling \cite{Bl}.
\vs{3mm}

\no{\Large\bf Acknowledgments}

\vs{3mm}

We would like to thank  E. J\"arvenp\"a\"a and I. Letawe
for their remarks. This work was
supported by NSF grant DMS-9205296, and by EC grant
CHRX-CT93-0411.

\end{document}